\documentclass[twoside,superscriptaddress,twocolumn,floatfix,a4paper,pra]{revtex4}
\usepackage{color}
\usepackage{graphicx}
\usepackage[utf8x]{inputenc}
\usepackage[T1]{fontenc}
\usepackage{siunitx}
\usepackage{amstext}
\usepackage{textgreek}
\usepackage{hyperref}
\usepackage{epstopdf}
\usepackage{amsfonts}
\usepackage{amsmath}
\usepackage{multirow}
\usepackage{tabularx}
\usepackage{filecontents}
\usepackage{float}

\newcommand\ket[1]{|#1\rangle}
\newcommand\bra[1]{\langle #1|}
\newcommand\oprod[2]{\ket{#1}\bra{#2}}
\newcommand\iprod[2]{\langle{#1}|{#2}\rangle}

\newcommand{\overbar}[1]{\mkern 1.5mu\overline{\mkern-1.5mu#1\mkern-1.5mu}\mkern 1.5mu}

\begin{document}

\title{Interplay between strong and weak measurement:\\ Comparison of three experimental approaches to weak value estimation}

\author{Jan Roik}
\email{jan.roik@upol.cz}
\affiliation{RCPTM, Joint Laboratory of Optics of Palacký University and Institute of Physics of Czech Academy of Sciences, 17. listopadu 12, 771 46 Olomouc, Czech Republic}

\author{Karel Lemr}
\email{k.lemr@upol.cz}
\affiliation{RCPTM, Joint Laboratory of Optics of Palacký University and Institute of Physics of Czech Academy of Sciences, 17. listopadu 12, 771 46 Olomouc, Czech Republic} 

\author{Antonín Černoch} \email{acernoch@fzu.cz}
\affiliation{Institute of Physics of the Czech Academy of Sciences, Joint Laboratory of Optics of PU and IP AS CR, 17. listopadu 50A, 772 07 Olomouc, Czech Republic}
   
\author{Karol Bartkiewicz} \email{bark@amu.edu.pl}
\affiliation{Faculty of Physics, Adam Mickiewicz University, PL-61-614 Pozna\'n, Poland}
\affiliation{RCPTM, Joint Laboratory of Optics of Palacký University and Institute of Physics of Czech Academy of Sciences, 17. listopadu 12, 771 46 Olomouc, Czech Republic}

\begin{abstract}
Weak values are traditionally obtained using a weak interaction between the measured system and a pointer state. It has, however, been pointed out that weak coupling can be replaced by a carefully tailored strong interaction. This paper provides a direct comparison of two strong interaction-based approaches (strong interaction accompanied by either a suitably prepared pointer state or quantum erasure) and the traditional weak interaction-based method. Presented theoretical derivations explicitly prove analytical equivalence of these approaches which was subsequently certified by an experiment implemented on the platform of linear optics. We find that strong-interaction-based measurements are experimentally less demanding on this platform. 
\end{abstract}

\date{\today}

\maketitle
\section{Introduction}
The concept of weak values was proposed by Aharonov, Albert and Vaidman in 1988 \cite{PhysRevLett.60.1351}. Although being a subject of some controversy \cite{PhysRevLett.62.2325,PhysRevLett.62.2326} and vivid scientific debate \cite{PhysRevA.88.046102,PhysRevA.94.052106,SOKOLOVSKI2017227,PhysRevA.94.032115,Alonso2015,PhysRevA.92.023829,PhysRevA.91.012103,PhysRevA.89.033825,PhysRevA.88.046103,PhysRevLett.111.240402,PhysRevA.97.026102,PhysRevA.95.066101,vaidman2017comment,Vaidman_2018a,PhysRevLett.62.2327,doi:10.1098/rsta.2016.0395,PhysRevA.93.036104,PhysRevA.93.036103,PhysRevA.93.017801,Vaidman_2018,vaidman2017comments,PhysRevLett.122.100405,PhysRevA.99.026103}, weak values allow acquiring information on a quantum state while causing arbitrarly small disturbance to the state. With the usage of weak interaction, we are able to measure a qubit without destroying it \cite{PhysRevLett.92.190402}, perform process tomography \cite{Kim2018}, directly measure the quantum wave function \cite{Lundeen2011}, observe trajectory in quantum systems in the semiclassical regime \cite{PhysRevLett.109.150407}, study the three-box paradox \cite{1751-8121-46-31-315307}, determine the past of photons passing through an interferometer \cite{PhysRevLett.111.240402} and even amplify the nonlinear effect of a single photon \cite{Hallaji2017}. Weak values have also being discussed in relation to the Hardy paradox \cite{PhysRevA.70.042102,AHARONOV2002130}.

It has been discovered that weak interaction is not essential for weak measurement \cite{PhysRevLett.94.220405,JOHANSEN2007374,RevModPhys.86.307} and the usage of strong interaction for weak value estimation is a hot current topic \cite{PhysRevA.98.042112,DENKMAYR2018339,vaidman2017comment}. Same claimed that strong interaction provides even better results \cite{PhysRevLett.116.040502} then direct weak measurement. However, it has been suggested that the interpretation of these results could be flawed \cite{vaidman2017comments}. Very recently, a study has been published showing that weak values can be analyzed independently of the strength of coupling between the pointer and the measured system \cite{Dziewior2881}. 

There are two major strategies for weak value estimation with strong interaction. First of them uses insensitive pointer, i.e. pointer prepared in a state that is close to an eigenstate of the strong interaction \cite{PhysRevA.93.032128,PhysRevA.93.062304,PhysRevA.91.052109,PhysRevLett.94.220405}. The second method applies quantum erasure where weak values are obtained by erasing results of a strong interaction from pointer state \cite{PhysRevLett.116.070404}. Although being theoretically proposed in 2016, this method has never been subjected to an experimental test. 

In this experimental study, we directly compare these two strong-interaction based approaches towards weak value estimation with the approach based on a genuine weak interaction used as a comparison baseline. The experiment is carried out on the platform of linear optics, more specifically we use the tunable controlled phase (c-phase) gate to achieve tunably strong interaction between the measured system and the quantum pointer. This paper, thus, contributes to the ongoing scientific discussion by lining up the result of the three strategies of weak value estimation on a single platform. This paper is organized as follows. In Sec.\ref{sec:Theoretical background} we provide the necessary theoretical background for our experiment. Sec.\ref{experiment} discusses the construction of linear-optical phase gate. In Sec.\ref{measurement}, we present our results along with the description of measured procedures. We conclude in Sec.\ref{conclusions}.
 
\section{Theoretical background} 
\label{sec:Theoretical background}
In this section, we outline a unifying framework to weak value estimation using three different approaches. We demonstrate this on an example of two interacting spin-$\tfrac{1}{2}$ particles (two qubits). For a single spin-$\tfrac{1}{2}$ particle we can express all possible unitary operations by means of generators of the associated unitary group, i.e., Pauli operators $X = \oprod{+}{+}-\oprod{-}{-},$  $Y = \oprod{L}{L}-\oprod{R}{R},$ and $Z = \oprod{0}{0}-\oprod{1}{1}.$ The eigenstates of $Z$ operator, labelled $\ket{0}$ and $\ket{1}$, correspond to various physical states depending on the particular two-level system. The eigenstates of the remaining operators are related in the following way: $\ket{\pm}=(\ket{0}\pm\ket{1})/\sqrt{2},$ $\ket{L}=(\ket{0}+i\ket{1})/\sqrt{2},$ and $\ket{R}=(\ket{0}-i\ket{1})/\sqrt{2}$. To facilitate the experimental comparison of the tested approaches, we have set in all instances the initial system state to 
\begin{equation}
|\psi_i\rangle= \cos \gamma |H\rangle + \sin \gamma |V\rangle,
\end{equation}
and after it interacts with the pointer we post-select it onto
\begin{equation}
|\psi_f\rangle=\tfrac{1}{\sqrt{2}}(|0\rangle+|1\rangle).
\end{equation}

Let us first review the traditional approach, i.e the typical formalism of weak value estimation by means of a weak interaction. For a detailed derivation, see Appendix. We define a general interaction Hamiltonian in the form of
\begin{equation}
H=A\otimes Z, 
\label{eq:ham}
\end{equation}
where $A$ and $Z$ act on the measured system and the probe respectively. We assume that the interaction between the measured system and the probe is sufficiently weak so that the evolution operator can be approximated as
\begin{equation}
U_w = e^{\frac{-itH}{\hbar}}\approx 1-\frac{itH}{\hbar}. 
\end{equation}
By considering post-selection on the final state of the weakly measured system
\begin{equation}
\begin{split}
&\oprod{\psi_{f}}{\psi_{f}} e^{\frac{-itH}{\hbar}} |\psi_{i}\rangle|\phi\rangle 
 \approx \iprod{\psi_{f}}{\psi_{i}} \ket{\psi_{f}}e^{\frac{-it A_w Z}{\hbar}}|\phi\rangle,
\end{split}
\end{equation}
weak value is obtained 
\begin{equation}
A_w =\frac{\langle\psi_{f}|A|\psi_{i}\rangle}{\langle\psi_{f}|\psi_{i}\rangle}.
\label{eq:weakvalue}
\end{equation}
In our experiment, weak interaction takes place in a c-phase gate and leads to a slight shift of the pointer state $|\phi\rangle \rightarrow |\phi' \rangle$. If we measure the pointer $\ket{\phi'}$ for observable $X,$ we obtain
\begin{equation}
\langle X \rangle_{\phi'}=1-|\tfrac{it}{\hbar}|^2|  A_w|^2.
\label{eq:X}
\end{equation}
In case of the tunable c-phase transformation the interaction Hamiltonian takes the form of 
\begin{equation}
H =  \tfrac{\varphi \hbar}{4t}(I-Z)\otimes Z,
\end{equation}
and hence the measured operator $A=Z-I$ (up to a coupling constant).

In the second regime (strong interaction and insensitive pointer), we assume that the interaction is strong and therefore the evolution operator is given as 
\begin{equation}
\label{eq:strong}
U_s = e^{\frac{-itH}{\hbar}}=\oprod{a_0}{a_0} \otimes I+\oprod{a_1}{a_1}\otimes Z,
\end{equation}
where $\ket{a_{0,1}}$ are eigenstates of observable $\overbar{A}=\oprod{a_0}{a_0}-\oprod{a_1}{a_1}$ of the measured system which is strongly coupled to observable $Z$ of the pointer. We conveniently parameterize the pointer state by two complex amplitudes $\alpha$ and $\beta$
\begin{equation}
|\phi\rangle=\alpha\ket{+}+\beta\ket{-}.
\end{equation}
In the situation when we set $\alpha=\beta,$ we deal with a completely insensitive pointer as $\langle Z \rangle_{\phi'}=1$ does not provide any information about the state of the measured system. Therefore to emulate weak coupling, we must chose the pointer state $|{\phi}'\rangle\approx|0\rangle,$ where
\begin{equation}
\alpha+\beta\approx\sqrt{2},\,\alpha-\beta=\delta,|\delta|\ll 1. 
\end{equation} 
In this case, if we measure the pointer for observable $X,$ we obtain
\begin{equation}
\langle X\rangle_{\phi'}=\mathrm{Re}(\delta \overbar{A}_w).
\label{eq:lan}
\end{equation} 
A detailed derivation presented in the Appendix also shows that in case of the c-phase gate $\overbar{A}=Z$ which relates this operator to the operator $A=\overbar{A}-I$ (up to coupling constant).

In the third and final regime (strong interaction followed by a quantum erasure), the pointer was prepared in the initial state $\ket{\phi} = \ket{+}$. In case of our experiment, the pointer state serves as the control in the c-phase gate set to strongest interaction (i.e., controlled-sign gate) and the measured system is its target. The choice of our particular gate (c-phase) determines the operator $\overbar{A}=Z$ of which the weak value is estimated. In this configuration, maximally entangled states can be created. Thus, this regime is also referred to as strong interaction. Next, the pointer state is measured in $\ket{\phi'}=\gamma\ket{0} + \delta\ket{1}$ and $\ket{\phi'_\perp}=\delta\ket{0} - \gamma\ket{1}$ basis, where $|\delta|\ll 1$ and $|\gamma| \approx 1.$ We measure the mean value of $Z= \ket{\phi'}\bra{\phi'}-\ket{\phi'_\perp}\bra{\phi'_\perp}$ we obtain  
\begin{equation}\label{eq:weak}
\langle Z \rangle_{\phi'} =\mathrm{Re}(4\delta \overbar{A}_w).
\end{equation}
Depending on a single-shot outcome outcome of this measurement, the system state is steered into $\left(\gamma+\delta\overbar{A}\right)|\psi_i\rangle$ or $\left(\gamma\overbar{A}-\delta\right)|\psi_i\rangle$. A feed-forward operation $\overbar{A}^{-1}$ is used in the second case to quasi-entirely revert the effect of this measurement on the measured system. For a detailed derivation, see Appendix. 

\section{Experimental setup}
\label{experiment}
 In our experiment, we encoded logical states $|0 \rangle$ and $|1 \rangle$ into horizontal $|H \rangle$ and vertical $|V \rangle$ polarization states of individual photons. Pairs of photons at \SI{710}{\nano\meter} were obtained in type-I spontaneous parametric down-conversion using a BBO ($ \beta- \text{BaB}_2\text{O}_3$) crystal. We accumulated the signal for \SI{100}{\second} to acquire about $400-3000$ two-photon coincidences depending on the particular setting.

\begin{figure}
		\begin{center}
		\includegraphics[scale=0.065]{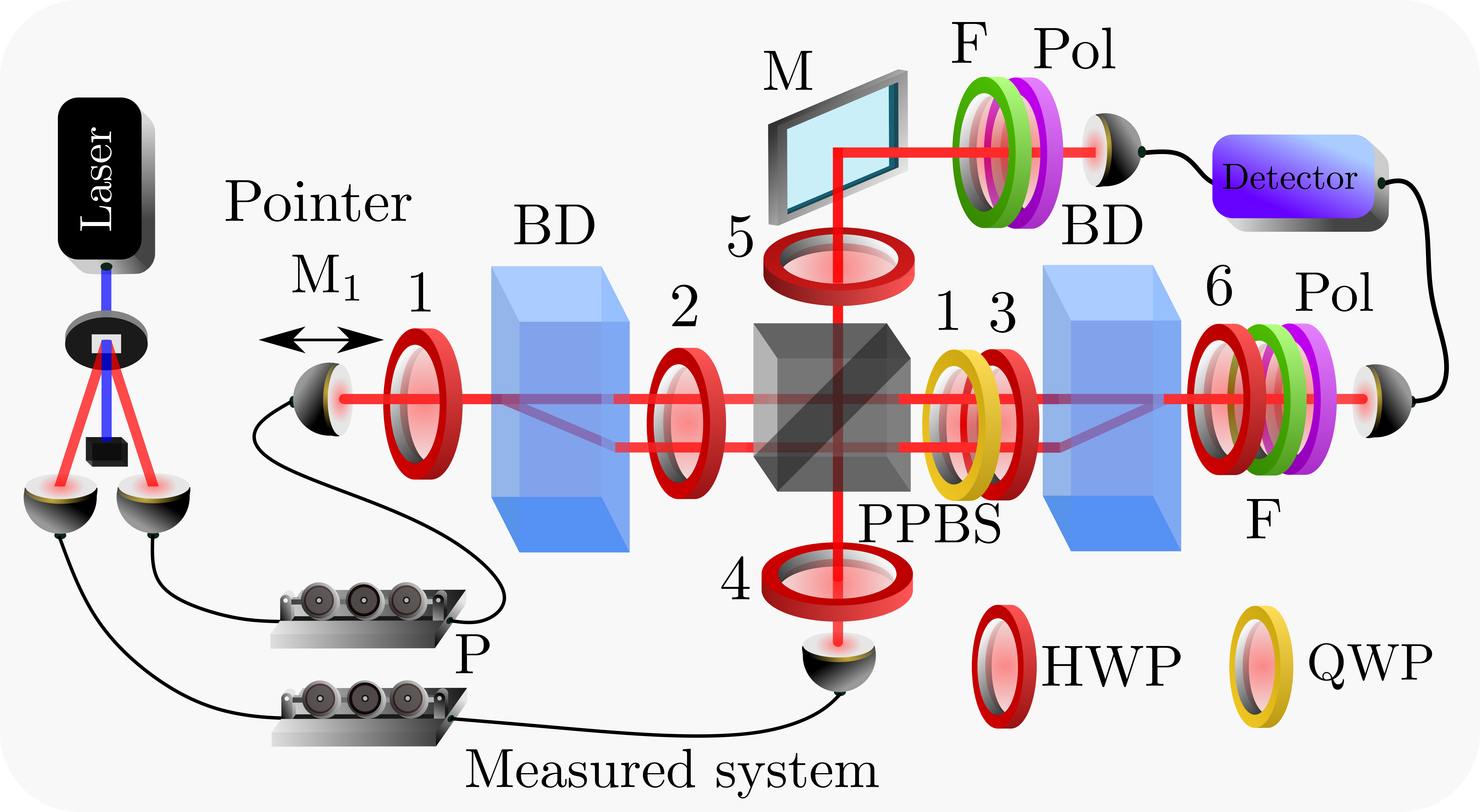}
		\caption{Experimental setup. P -- polarization controller, M -- mirror, BD -- beam displacer, F -- spectral filter (\SI{5}{\nano\meter} wide) $\text{M}_1$ -- motorized translation, PPBS -- partially polarizing beam splitter, HWP -- half-wave plate, QWP -- quarter-wave plate, Pol -- polarizer.}
		\label{Fig2}		
		\end{center}
\end{figure}

We have constructed a tunable linear-optical c-phase gate based on the design by Lanyon {\it et al.} \cite{Lanyon2008}. This experimental setup (see Fig.~\ref{Fig2}) was chosen for its stability and easily adjustable phase shift $\varphi$. Note that this gate is experimentally much less demanding than the optimal tunable c-phase gate consisting of three nested interferometers \cite{PhysRevLett.106.013602}. 
 This c-phase gate consists of only one interferometer formed by two beam displacers with a partially polarizing beam splitter (PPBS) acting in one of it's arms. The PPBS is manufactured to have $100\%$ transmissivity for horizontally polarized light and $33\%$ transmissivity for vertically polarized photons. As a result, it implements a c-phase gate with $ \varphi= \pi$ on impinging photons due to the Hong-Ou Mandel interference \cite{PhysRevLett.95.210505}. In this special case, the gate is often called controlled sign (c-sign) gate. The idea to use a special PPBS to implement c-sign gate can be traced back to Ref.\cite{PhysRevLett.95.210505}. To obtain phase shifts other than $ \varphi= \pi$, one needs to make use of an ancillary mode provided by the interferometer. 
 
The working principle of this tunable c-phase gate can be explained by describing the role of individual components in the setup. First, a beam displacer (BD) spatially splits polarization modes of the pointer photon depending on its state. If the photon is encoded in the logical $|1\rangle$ state (vertical polarization $|V\rangle$) it travels through the upper arm where it impinges on the PPBS. Otherwise, the pointer photon in the logical $|0\rangle$ state (horizontal polarization $|H\rangle$) passes through the interferometer without interacting on the PPBS. On the PPBS, the pointer photon can interact with the photon that represents the measured system. As mentioned above, this interaction implements the c-sign gate. The combination of c-sign gate (made by PPBS) together with the quarter-wave plate ($\text{QWP}_1$) transforms the pointer to the $|R\rangle$ (right-handed circular) or $|L \rangle$ (left-handed circular) polarization state depending on the signal photon being in $|H \rangle$ or $|V \rangle$ state respectively. Next, a half-wave plate introduces a phase shift $ \frac{\varphi}{2}$ and $-\frac{\varphi}{2}$ to the $|R\rangle$ and $|L \rangle$ states respectively, with the value of $\varphi$ depending on its rotation. Finally, both interferometer arms recombine on the second BD.

As outlined above, if the pointer is prepared in the $|H\rangle$ state, it travels through the setup by the lower interferometer arm (see setup  in Fig.\ref{Fig2}) and can not interact with the measured system. The gate only acts non-trivially if the pointer is prepared vertically polarized. Table \ref{Tab1} describes transformation of the joint pointer and measured system states as they pass through the setup in two non-trivial cases.
\begin{table}
\caption{Output two-mode states after being transformed by consecutive components of the c-phase gate. The modes  denote polarization states of pointer and measured photons respectively, $|+_{60}\rangle =\frac{1}{2}|H\rangle +\frac{3}{2}|V\rangle$, i.e.  linear polarization rotated by \ang{60} with respect to horizontal polarization \cite{PhysRevA.87.062333}.  Post-selection on coincident detection of the photons in their respective output ports is assumed.}
\begin{ruledtabular}
\begin{tabular}{lcc}

Component                        & Case $|VH\rangle$ 						& Case $|VV\rangle$ \\ \hline
First BD                         & $|VH\rangle$     						& $|VV\rangle$ \\
$\text{HWP}_2$ at \ang{30}       & $|+_{60}H\rangle$    				& $|+_{60}V\rangle$ \\
PPBS                             & $|+H\rangle$      						& $|-V\rangle$ \\
$\text{QWP}_1$ at \ang{0}	    & $|RH\rangle$      						& $|LV\rangle$ \\
$\text{HWP}_3$ as needed         & $e^\frac{-i\varphi}{2}|LH\rangle$      & $e^\frac{i\varphi}{2}|RV\rangle$ \\
Second BD                        & $e^\frac{-i\varphi}{2}|HH\rangle$      & $e^\frac{i\varphi}{2}|HV\rangle$ \\
$\text{HWP}_6$   at \ang{45}     & $e^\frac{-i\varphi}{2}|VH\rangle$      & $e^\frac{i\varphi}{2}|VV\rangle$ \\
\end{tabular}
\end{ruledtabular}
\label{Tab1}	
\end{table}

\section{Measurement and results}
\label{measurement}
\subsection{Weak measurement by weak interaction}

\begin{figure}
		\begin{center}
		\includegraphics{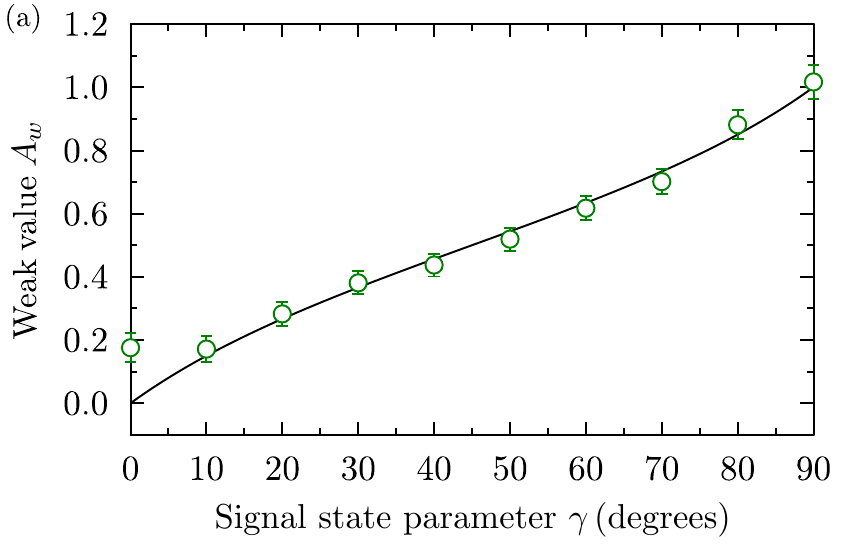}
		\includegraphics{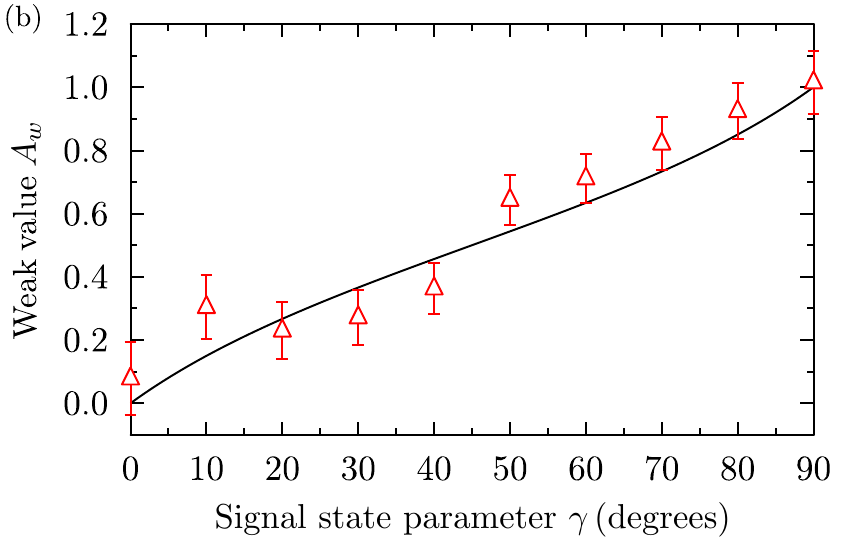}
		\includegraphics{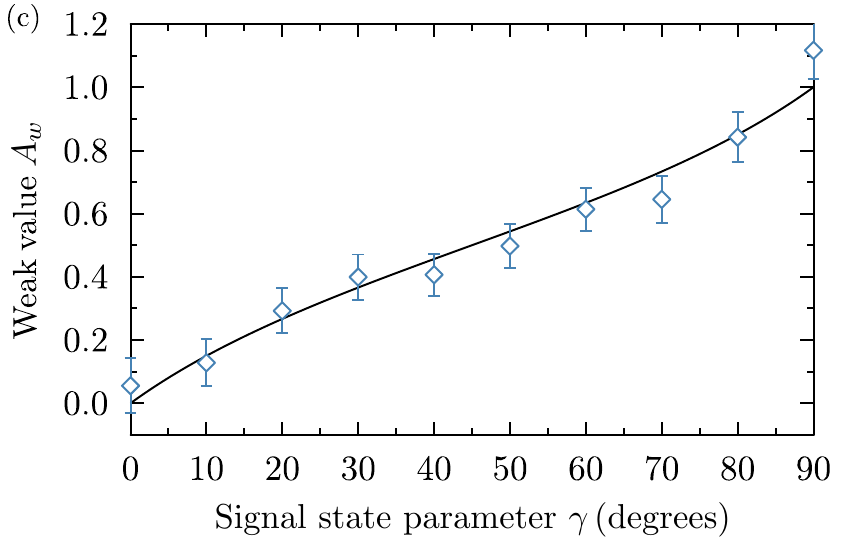}
		\caption{Comparison of fit of theoretical dependence on experimental data (solid lines) and experimentally measured weak values (markers) using the approaches: (a) weak value estimation by weak interaction, (b) weak value estimation by strong interaction and insensitive pointer, (c) weak value estimation by strong interaction and quantum erasure. Weak values are depicted as functions of the signal state parameter $\gamma$.}
		\label{fg:data}		
		\end{center}
\end{figure}

The first implemented regime corresponds to the traditional weak value estimation (weak value estimation by weak interaction) and provides data for comparison to the other regimes of weak value estimation. In this regime, we tuned the gate to a small but still measurable phase shift $ \varphi$. The input signal was prepared in the state 
\begin{equation}
|\psi_i\rangle= \cos \gamma |H\rangle + \sin \gamma |V\rangle,
\label{eq:deffit}
\end{equation}
parameterized by the angle $\gamma$. For practical reasons, we have selected final post-selection onto state $| \psi_f \rangle= \frac{1}{ \sqrt{2}}$ $(|H \rangle +|V \rangle)$.
The initial pointer state was set to $|\phi\rangle=\frac{1}{\sqrt{2}}(|H\rangle+|V\rangle)$ prepared by $\text{HWP}_1$. For these settings Eq.~(\ref{eq:weakvalue}) gives the theoretical prediction for the weak value as a function of state parameter $\gamma$ 
\begin{equation}
A_w  = \frac{1}{\cot \gamma+1}.
\label{eq:coth}
\end{equation}
The pointer state after interaction with the signal $|\phi'\rangle$ was measured in the 
$\frac{1}{ \sqrt{2}}$ $(|H \rangle \pm|V \rangle)$ basis and the value of $\langle X \rangle_{\phi'}$ was calculated. This value was then translated to $ A_w $ using Eq.~(\ref{eq:X}) and fitting of the parameter $\varphi\approx0.18\pi$. $\varphi$ was set by suitable rotation of $\text{HWP}_3$. Results of this approach are visualized in Fig.~\ref{fg:data}a. Our data also reveals that typical fidelity of the output measured system state before its post-selection with respect to its initial state $|\psi_i\rangle$ is about 96 \%. This confirms that the system was weakly perturbed by the interaction with the pointer.

\subsection{Weak value estimation by strong interaction and insensitive pointer}

The second regime was implemented by using the PPBS as a c-sign gate with $\varphi=\pi$. The lower interferometer arm was blocked. All the pointer photons were sent to the upper arm and second BD was used for polarization projection during the measurement. We used the same set of input signal $|\psi_i\rangle$ and  final $|\psi_f\rangle$ states as in previous regime. The pointer initial state was prepared by the $\text{HWP}_2$ set to \ang{3} resulting in a state close to $|0\rangle$. Again, the pointer was measured in the $\frac{1}{\sqrt{2}}(|H\rangle\pm|V\rangle)$ basis using $\text{HWP}_3$. Weak value $ \overbar{A}_w $ was obtained using Eq. (\ref{eq:lan})
and fitting of parameter $\delta \approx 0.21$. Results of this approach are visualized in Fig.~\ref{fg:data}b.

\subsection{Weak value estimation by strong interaction and quantum erasure}

The third, and final, regime was implemented again by using only the upper interferometer arm and PPBS. In this case, the pointer was prepared in the $|\phi \rangle = \frac{1}{\sqrt{2}}(|H\rangle+|V\rangle)$ state by $\text{HWP}_2$ and projected in a basis close to $|H\rangle/|V\rangle$ using $\text{HWP}_3$. The feed-forward required in this regime was simulated by projecting the measured system onto $\frac{1}{\sqrt{2}}(|H\rangle+|V\rangle)$ and $\frac{1}{\sqrt{2}}\left(|H\rangle-|V\rangle\right)$ when the pointer was projected onto states close to $|H\rangle$ and $|V\rangle$ respectively. Weak value $ \overbar{A}_w $ was obtained using Eq. (\ref{eq:weak}) and fitting of parameter $\delta \approx 0.08$. Results of this approach are visualized in Fig.~\ref{fg:data}c.

In Fig.~\ref{fg:data}, we compare experimental results obtained using all methods with the theoretical prediction (\ref{eq:coth}). Note that in case of the two strong interaction-based approaches $ \overbar{A}_w $ was translated to $ A_w $ using $ A_w = \frac {(1-  \overbar{A}_w )}{2}$ as explained in the Appendix. Our data reasonably overlaps with theoretical derivations proving the working principle of all approaches.

\section{Conclusions}
\label{conclusions}

We have experimentally compared three approaches to weak value estimation. Two of these approaches are based on strong interaction between the pointer and measured system, the third uses the original idea of weak coupling. We have established that all these approaches lead to identical results up to experimental uncertainties. This is in accordance with our theoretical framework unifying these approaches.

In our particular implementation, adjusting the setup for the quantum erasure seemed to be slightly less demanding in comparison to the approach using an insensitive pointer state. Both these approaches are considerably less demanding to implement on our platform of choice because they do not necessitate tunable c-phase gate. As such, they can be implemented by Hong-Ou-Mandel-type interference alone accompanied by suitable single photon polarization manipulation. Implementing a weak c-phase gate interaction (i.e. small values of phase shift $\varphi$) is known to require ancillary modes and thus an second-order interferometric setup on top of the above mentioned Hong-Ou-Mandel interference. Note however, that these conclusions are platform and gate-type specific. It is also worth noting that c-phase operation can be achieved nearly deterministically if set to phase shifts close to zero. This is definitely a benefit of the traditional approach to weak value estimation, at least on the platform of linear optics. Linear-optical setup implementing this quasi-deterministic operation is however significantly more complex.

Our quantum erasure approach bear some similarities to \cite{PhysRevA.91.052109}, but adds a feed-back to erase the influence of strong coupling.

%

\section*{Acknowledgement}
Authors thank Cesnet for providing data management services. Authors acknowledge
financial support by the Czech Science Foundation under the project No. 17-10003S. KB also acknowledges the financial support of the Polish National Science Center under grant No. DEC-2015/19/B/ST2/01999. The authors also acknowledge the project No. CZ.02.1.01./0.0/0.0/16\textunderscore 019/0000754 of the Ministry of Education, Youth and Sports of the Czech Republic. JR also acknowledges the Palacky University internal grant No. IGA-PrF-2019-008.

\section*{Appendix: Theoretical framework}
\label{th model}
This appendix provides detailed derivation of the unifying framework to weak value estimation using three different approaches to weak value estimation. We demonstrate this on an example of two interacting spin-$\tfrac{1}{2}$ particles (two qubits). For a single spin-$\tfrac{1}{2}$ particle we can express all possible unitary operations by means of generators of the associated unitary group, i.e., Pauli operators $X = \oprod{+}{+}-\oprod{-}{-},$  $Y = \oprod{L}{L}-\oprod{R}{R},$ and $Z = \oprod{0}{0}-\oprod{1}{1}.$ The eigenstates of $Z$ operator, labelled $\ket{0}$ and $\ket{1}$, correspond to various physical states depending on the particular two-level system. The eigenstates of the remaining operators are related in the following way: $\ket{\pm}=(\ket{0}\pm\ket{1})/\sqrt{2},$ $\ket{L}=(\ket{0}+i\ket{1})/\sqrt{2},$ and $\ket{R}=(\ket{0}-i\ket{1})/\sqrt{2}$.

\subsection{Weak measurement by weak interaction}
\begin{figure}
		\begin{center}
		\includegraphics[scale=0.41]{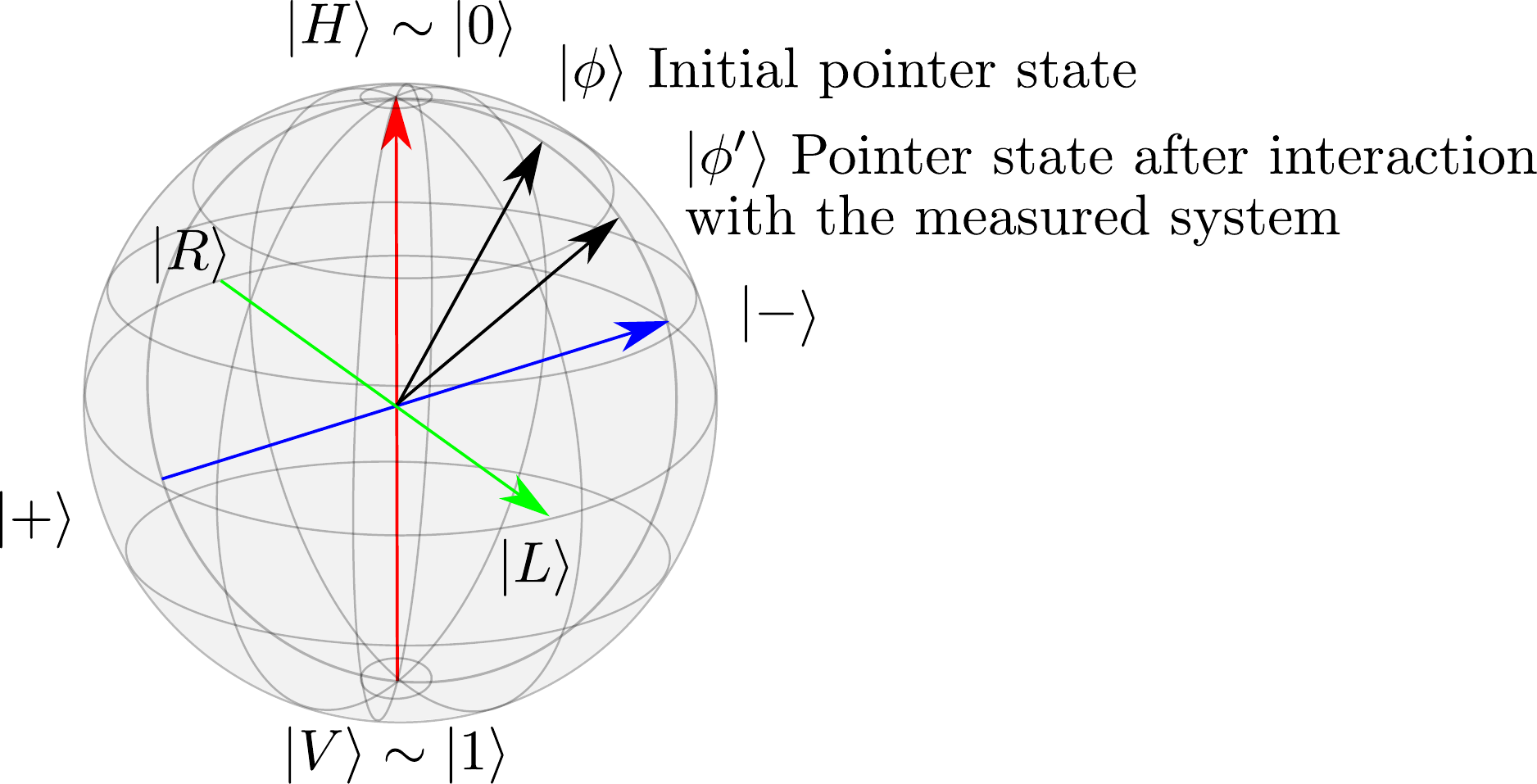}
		\caption{Rotation of the pointer state due to weak interaction.}
		\label{Bloch}		
		\end{center}
\end{figure}

We start by reviewing the typical formalism of weak value estimation by means of a weak interaction. Let us assume that $|\Psi_{t}\rangle$ is the overall input state 
\begin{equation}
|\Psi_{t}\rangle=|\psi_{i}\rangle|\phi\rangle
\end{equation}
consisting of the measured system $|\psi_{i}\rangle$ and a pointer state $|\phi\rangle$. Let the interaction Hamiltonian between these two subsystems be specified as
\begin{equation}
H=A\otimes Z, 
\label{eq:ham1}
\end{equation}
where $A$ and $Z$ act on the measured system and the probe respectively. The evolution induced by time-independent Hamiltonian can be represented as $e^{\frac{-itH}{\hbar}}$. We now assume that the interaction between the measured system and the probe is sufficiently weak so that the evolution operator can be approximated as
\begin{equation}
U_w = e^{\frac{-itH}{\hbar}}\approx 1-\frac{itH}{\hbar}. 
\end{equation}
Let us consider post-selection on the final state of the weakly measured system
\begin{equation}
\begin{split}
&\oprod{\psi_{f}}{\psi_{f}} e^{\frac{-itH}{\hbar}} |\psi_{i}\rangle|\phi\rangle 
 \approx \oprod{\psi_{f}}{\psi_{f}}\left( 1-\frac{itH}{\hbar}\right)|\psi_{i}\rangle|\phi\rangle \\
&  = \oprod{\psi_{f}}{\psi_{f}}\psi_{i}\rangle\left(1-\frac{it}{\hbar}\frac{\langle\psi_{f}|A|\psi_{i}\rangle}{\langle\psi_{f}|\psi_{i}\rangle}Z\right)|\phi \rangle \\
&  = \iprod{\psi_{f}}{\psi_{i}} \ket{\psi_{f}}e^{\frac{-it A_w Z}{\hbar}}|\phi\rangle.
\end{split}
\end{equation}
Here, 
\begin{equation}
 A_w=\frac{\langle\psi_{f}|A|\psi_{i}\rangle}{\langle\psi_{f}|\psi_{i}\rangle}
\label{eq:weakvalue1}
\end{equation}
represents the weak value and $e^{\frac{-it A_w Z}{\hbar}}|\phi\rangle $ is the output pointer state. If we take   
\begin{equation}
|\phi\rangle=|+\rangle=\tfrac{1}{\sqrt{2}}(|0\rangle+|1\rangle),
\label{eq:plus}
\end{equation}
as the initial pointer state, after some time we obtain
 \begin{equation}
\begin{split}
&|\phi\rangle  \rightarrow  |{\phi}'\rangle=(1-\tfrac{it}{\hbar} A_w  Z)|\phi\rangle,
\end{split}
\end{equation}
which leads to 
\begin{equation}\label{eq:WIpointer}
|{\phi}'\rangle =\ket{+}-\tfrac{it}{\hbar} A_w \ket{-}.
\end{equation}
The final pointer state is weakly shifted with respect to its initial state (see Fig.\ref{Bloch}). The magnitude of the shift is dependent on the  force of interaction. 
If we measure the pointer $\ket{\phi'}$ for observable $X,$ we obtain
\begin{equation}
\langle X \rangle_{\phi'}=1-|\tfrac{it}{\hbar}|^2|  A_w|^2.
\label{eq:X1}
\end{equation}
Alternatively, if we measure observable $Z,$ the outcome is
\begin{equation}
\langle Z\rangle_{\phi'}=2i\,\mathrm{Im}\left(\tfrac{it}{\hbar} A_w\right) = \mathrm{Re}\left(\tfrac{2t}{\hbar} A_w\right).
\label{eq:lan2}
\end{equation}
Similarly, the real part of $\tfrac{it}{\hbar} A_w$ can be estimated by measuring $Y$. All the above-listed equations do not depend on the size of the investigated subsystems. In particular, the pointer can be an arbitrarily large system instead of a qubit and $Z$ operator will be a high-dimensional counterpart of the respective Pauli's operator.

\subsection{Weak measurement by strong interaction and insensitive pointer} 
\label{sec:insensitive}
Here, we demonstrate our first strategy to weak value estimation even with strong interaction (i.e., controlled-sign gate), given as
\begin{equation}
\label{eq:strong1}
U_s = e^{\frac{-itH}{\hbar}}=\oprod{a_0}{a_0} \otimes I+\oprod{a_1}{a_1}\otimes Z,
\end{equation}
where $\ket{a_{0,1}}$ are eigenstates of observable $\overbar{A}=\oprod{a_0}{a_0}-\oprod{a_1}{a_1}$ of the measured system which is strongly coupled to observable $Z$ of the pointer. These states form a basis for a spin-$\tfrac{1}{2}$ system, i.e., $I=\oprod{a_0}{a_0}+\oprod{a_1}{a_1}$. Similarly to the case of weak interaction, we apply the unitary operator $U_s$ to the measured system and probe states and subsequently post-select the measured system onto $|\psi_f\rangle$
\begin{equation}
\begin{split}
&\oprod{\psi_f}{\psi_f} e^{\frac{-itH}{\hbar}} \ket{\psi_{i}}\ket{\phi}\\
&=\ket{\psi_f}\left(\frac{\langle\psi_f| a_0\rangle\langle a_0|\psi_{i}\rangle}{\langle\psi_f|\psi_{i}\rangle}I+\frac{\langle\psi_f| a_1\rangle\langle a_1|\psi_{i}\rangle}{\langle\psi_f|\psi_{i}\rangle}Z\right)|\phi\rangle.
\end{split}
\label{eq:psif}
\end{equation}
For conveninece, let us introduce operator $B$ defined as
$B = \bra{\psi_f} U_s \ket{\psi_{i}}.$
One can show by direct calculations that
\begin{equation}
\begin{split}
&B=(\langle\psi_f| \overbar{A} |\psi_{i}\rangle+\langle\psi_f |a_1\rangle\langle a_1|\psi_{i}\rangle)\frac{I}{\langle\psi_f|\psi_{i}\rangle}\\
&-(\langle\psi_f| \overbar{A} |\psi_{i}\rangle-\langle\psi_f| a_0\rangle\langle a_0|\psi_{i}\rangle)\frac{Z}{\langle\psi_f|\psi_{i}\rangle}.
\end{split}
\label{eq:A}
\end{equation}
Equation (\ref{eq:A}) can be further simplified to obtain
\begin{equation}
2B=(1+ \overbar{A}_w)I+(1- \overbar{A}_w)Z,
\end{equation}
where 
\begin{equation}
 \overbar{A}_w=\frac{\langle\psi_f|\overbar{A}|\psi_{i}\rangle}{\langle\psi_f|\psi_{i}\rangle}
\label{eq:WeakEr}
\end{equation} 
is the weak value. We now use this result to rewrite Eq.~(\ref{eq:psif}) as
\begin{equation}
\ket{\psi_f}B|\phi\rangle = \tfrac{1}{2}\ket{\psi_f}
[({1+ \overbar{A}_w}) I+(1- \overbar{A}_w)Z)]|\phi\rangle.
\end{equation}
The shift in ancilla state is then 
\begin{equation}
|{\phi}'\rangle=\tfrac{1}{2}[(1+ \overbar{A}_w)I+({1- \overbar{A}_w}Z)]|\phi\rangle.
\end{equation}
Let us parametrise the initial probe state by amplitudes $\alpha$ and $\beta$
\begin{equation}
|\phi\rangle=\alpha\ket{+}+\beta\ket{-}.
\end{equation}
Expressing the shifted probe state $|{\phi}'\rangle$ in terms of eigenstates of the observable $Z$ yields
\begin{equation}
|{\phi}'\rangle=\frac{\alpha+\beta}{\sqrt{2}}\ket{0}-\frac{\alpha-\beta}{\sqrt{2}} \overbar{A}_w \ket{1}.
\end{equation} 
In the situation when we set $\alpha=\beta,$ we deal with a completely insensitive pointer as $\langle Z \rangle_{\phi'}=1$ does not provide any information about $ \overbar{A}_w$. On the other hand, if we set $\alpha=1$ and $\beta=0$, we obtain
\begin{equation}
|{\phi}'\rangle  =\tfrac{1}{2}[(1- \overbar{A}_w)|+\rangle-(1+ \overbar{A}_w)|-\rangle],
\end{equation}
which leads to a strong measurement
\begin{equation}
\langle Z\rangle_{\phi'}=\tfrac{1}{2}(1-| \overbar{A}_w |^2).
\label{eq:la}
\end{equation}
Now, let us consider measurement of $X$ on probe state $|{\phi}'\rangle$. In case of a completely insensitive pointer (i.e., $\alpha=\beta$ and $|{\phi}'\rangle=|0\rangle$) we observe $\langle X\rangle_{\phi'}=0$. If $\alpha=1,\, \beta=0,$  and we use strong measurement the expected value of $X$ in state $\ket{\phi'}$ is
\begin{equation}
\langle X \rangle_{{\phi'}}=\tfrac{1}{4}|(1- \overbar{A}_w)|^2-\tfrac{1}{4}|(1+ \overbar{A}_w)|^2= \mathrm{Re} \overbar{A}_w.
\end{equation}
Finally, let us consider pointer state $|{\phi}'\rangle\approx|0\rangle,$ where
\begin{equation}
\alpha+\beta\approx\sqrt{2},\,\alpha-\beta=\delta,|\delta|\ll 1, 
\label{eq:alpha}
\end{equation} 
which is transformed due to strong interaction into
\begin{equation}\label{eq:SIpointer}
|{\phi}'\rangle=|0\rangle+\tfrac{\delta}{2} \overbar{A}_w|1\rangle.
\end{equation}
By slightly deviating the pointer from $\ket{0}$ state we made it weakly sensitive to the interacting system.

In this case, if we measure the pointer for observable $Z,$ we obtain
\begin{equation}
\langle Z \rangle_{\phi'}=1-|\tfrac{\delta}{2}|^2|  \overbar{A}_w|^2.
\end{equation}
Alternatively, if we measure observable $X,$ the outcome is
\begin{equation}
\langle X\rangle_{\phi'}=\mathrm{Re}(\delta \overbar{A}_w).
\label{eq:lan3}
\end{equation}
Similarly, the imaginary part of $\delta \overbar{A}_w$ can be estimated by measuring $Y$. 
Note that in the above derivations we consider a coupled system of two qubits, however, the results are also valid  in the general case, where (i) the pointer is not necessarily a spin-$\tfrac{1}{2}$ system, (ii) $\bar{A}=\Pi_0-\Pi_1$ is a multidimensional operator, and its positive valued components satisfy the completeness relation $\Pi_0+\Pi_1=I$.

\subsection{Strong interaction and subsequent erasure}
\label{sec:erasure}

Consider a pointer prepared in state $\ket{\phi} = \ket{+},$  and system in preselected state $\ket{\psi_i}=\alpha\ket{0} + \beta\ket{1}.$ Let the pointer be in control mode of a c-phase gate set to 
strongest interaction (i.e., controlled-sign gate) and let the system be in the target mode. The choice of this particular gate determines the operator $\overbar{A}=Z$ of which the weak value is estimated. In this configuration, maximally entangled states can be created. Thus, this regime is referred as strong interaction. Next, the pointer state is measured in $\ket{\phi'}=\gamma\ket{0} + \delta\ket{1}$ and $\ket{\phi'_\perp}=\delta\ket{0} - \gamma\ket{1}$ basis, where $|\delta| \ll 1$ and $|\gamma| \approx 1 .$ Now, the joint state of the control and target modes reads as
\begin{equation}
\ket{\Psi_t} = \tfrac{1}{\sqrt{2}}[(\gamma +\delta I\otimes \overbar{A})\ket{\phi'}\ket{\psi_i} + (\gamma I\otimes \overbar{A}-\delta) \ket{\phi'_\perp} \overbar{A}\ket{\psi_i}].
\label{strongerasure}
\end{equation}
By measuring the pointer we obtain
\begin{eqnarray}
\langle \phi'|\Psi_t\rangle&=& \tfrac{1}{\sqrt{2}}(\gamma+\delta  \overbar{A})\ket{\psi_i},\\
\langle \phi'_\perp|\Psi_t\rangle&=& \tfrac{1}{\sqrt{2}}(\gamma \overbar{A}-\delta)\ket{\psi_i}.
\end{eqnarray} 
It is obvious that the state of the system is perturbed  and its either $\langle \phi'|\Psi\rangle$ or $\langle \phi'_\perp|\Psi\rangle,$ depending on the measurement outcome of the pointer. If we use $\overbar{A}^{-1}$ operation on the target mode conditioned on finding the pointer in  $\ket{\phi'_\perp}$ state, we always find the system (up to a global phase factor) in a weakly perturbed state 
$$\ket{\psi'_i}=(\gamma+\delta  \overbar{A})\ket{\psi_i}.$$ 
Let us consider what happens when we measure operator $Z= \ket{\phi'}\bra{\phi'}-\ket{\phi'_\perp}\bra{\phi'_\perp}$ on the pointer state. 
Similarly, we can construct $X$ and $Y$ observables, corresponding to Pauli operators.
In this case the system state is postselected as usual. The pointer state is detected in $\ket{\phi'}$ with probability
\begin{equation}
|\langle \psi_f|\psi'_i \rangle|^2 \approx |\langle \psi_f|\psi_i \rangle|^2[1+2\mathrm{Re}(\delta \overbar{A}_w)],
\label{pointerprobability}
\end{equation}
whereas for projecting the pointer on $\ket{\phi'_\perp}$ and performing feed-forward $\overbar{A}^{-1}$ operation on the strongly perturbed system (here $\overbar{A}=\overbar{A}^{-1}$) we obtain 
\begin{equation}
|\langle \psi_f|\psi'_i \rangle|^2 \approx |\langle \psi_f|\psi_i \rangle|^2[1-2\mathrm{Re}(\delta \overbar{A}_w)].
\end{equation}
Thus, the mean value of $\langle Z \rangle_{\phi'}$ observed on the pointer for preselected and postselected states $\ket{\psi'_i}$ and $\bra{\psi_f},$ respectively, is 
\begin{equation}\label{eq:weak1}
\langle Z \rangle_{\phi'} =\mathrm{Re}(4\delta \overbar{A}_w).
\end{equation}
We can observe the imaginary part of the weak value if $Y$ operator is used instead of $Z.$
For the clarity of presentation the above-described derivations focus on a coupled pair of qubits. However, the results also hold when (i) $\bar{A}$ is non-singular (unitary), (ii) target mode is an arbitrarily large system, (iii) the pointer can be an arbitrarily large system that can be projected onto either $\ket{\phi}$ or $\ket{\phi_\perp},$ and their superpositions corresponding to eigenstates of high-dimensional counterparts of Pauli operators.

\subsection{Operational equivalence}
When we look at  Eqs.~(\ref{eq:lan2}), (\ref{eq:lan3}) and (\ref{eq:weak1}), we can conclude that probes carry information about weak values of the measured system. By measuring the deviation of the respective pointer from its initial position we can estimate these weak values. This proves that the approaches are operationally equivalent and the values of parameters $\delta$ and $it/\hbar$ play the roles of strength controllers in their respective scenarios. Weak value estimation was proposed to acquire small amount of information about the measured system while minimizing inflicted disturbance. Similarly, by using a pointer state insensitive to the interaction, we do not perturb the measured system. It can easily be seen, that for $\delta\rightarrow 0$, the probe state becomes the eigenstate of the $Z$ operator. Inserting this eigenstate relation into Eq. (\ref{eq:strong1}) factorizes the action onto the measured system and the probe. As a result, regardless of the measurement outcome on the probe, the measured system remains unperturbed. In the same manner, quantum erasing allows to tune the trade-off between the information gained about the system and the damage caused to it. One can verify that in case of complete erasure $\delta\rightarrow 0$ and correct feed-forward, the measured system is not changed. 
 
We demonstrate the above described theoretical concept on an example of a tunable controlled-phase (c-phase) gate. This gate transforms a pair of qubits according to a prescription 
\begin{equation}
|mn \rangle \rightarrow e^{ i \varphi \delta_{m1} \delta_{n1}}|mn \rangle, 
\end{equation}
where $m,n \in \{ 0,1 \}$ stand for logical states of the qubits, $\varphi$ denotes the  introduced phase shift and $\delta$ is the Kronecker's delta. By setting the value $\varphi$ we change the strength of the interaction between the two qubits from infinitesimally weak $\varphi \rightarrow 0$ to maximally strong $\varphi = \pi$.
Note that, if we set $A=\frac{I-Z}{2}$ in Eq.~(\ref{eq:ham}), then the weak interaction (up to the coupling constant)  corresponds to the controlled-phase gate. The Hamiltonian of which reads as
\begin{equation}
H =  \tfrac{\varphi \hbar}{4t}(I-Z)\otimes(I-Z)
\end{equation}
and can be divided into two terms  $H=H_1 + H_0$ describing the interaction
\begin{equation}
H_1 = \tfrac{\varphi \hbar}{4t} (Z-I)\otimes Z 
\end{equation}
and free evolution of the pointer
\begin{equation}
H_0 = \tfrac{\varphi \hbar}{4t}(I- Z)\otimes I.
\end{equation}
Similarly, the strong interaction in Eqs.~(\ref{eq:strong}) and (\ref{strongerasure}) can also be implemented using a c-phase gate, this time with $\varphi=\pi$. In this case, the unitary operator $U_s$ reads 
\begin{equation}
\label{operator}
U_s=|0\rangle\langle 0|\otimes I+|1\rangle\langle 1| \otimes Z
\end{equation}
and hence $\overbar{A}=Z$.
 Thus, while comparing  weak value estimation by weak interaction and by strong interaction together with either insensitive pointer or quantum erasure we simply fix the phase  $\varphi$ of the weak interaction to match  $\delta$. Moreover, we need to take into account free evolution of the composite system described by $H_0$ which will result in independent phase shifts between eigenstates of $Z$ for both system and pointer. Weak values $ A_w$ [Eq.~(\ref{eq:weakvalue})] and $\overbar{A}_w $ [Eqs.~(\ref{eq:WeakEr}), (\ref{pointerprobability})] obtained using weak and strong interactions respectively can be directly compared using a linear relation between operators A and $\overbar{A}$ 
 \begin{equation}
 A=\frac{I-\overbar{A}}{2}.
\end{equation}  
Which translates into 
\begin{equation}
A_w  = \frac{1- \overbar{A}_w}{2}.
\end{equation}


\begin{thebibliography}{49}
\expandafter\ifx\csname natexlab\endcsname\relax\def\natexlab#1{#1}\fi
\expandafter\ifx\csname bibnamefont\endcsname\relax
  \def\bibnamefont#1{#1}\fi
\expandafter\ifx\csname bibfnamefont\endcsname\relax
  \def\bibfnamefont#1{#1}\fi
\expandafter\ifx\csname citenamefont\endcsname\relax
  \def\citenamefont#1{#1}\fi
\expandafter\ifx\csname url\endcsname\relax
  \def\url#1{\texttt{#1}}\fi
\expandafter\ifx\csname urlprefix\endcsname\relax\def\urlprefix{URL }\fi
\providecommand{\bibinfo}[2]{#2}
\providecommand{\eprint}[2][]{\url{#2}}

\bibitem[{\citenamefont{Aharonov et~al.}(1988)\citenamefont{Aharonov, Albert,
  and Vaidman}}]{PhysRevLett.60.1351}
\bibinfo{author}{\bibfnamefont{Y.}~\bibnamefont{Aharonov}},
  \bibinfo{author}{\bibfnamefont{D.~Z.} \bibnamefont{Albert}},
  \bibnamefont{and} \bibinfo{author}{\bibfnamefont{L.}~\bibnamefont{Vaidman}},
  \bibinfo{journal}{Phys. Rev. Lett.} \textbf{\bibinfo{volume}{60}},
  \bibinfo{pages}{1351} (\bibinfo{year}{1988}),
  \urlprefix\url{https://link.aps.org/doi/10.1103/PhysRevLett.60.1351}.

\bibitem[{\citenamefont{Leggett}(1989)}]{PhysRevLett.62.2325}
\bibinfo{author}{\bibfnamefont{A.~J.} \bibnamefont{Leggett}},
  \bibinfo{journal}{Phys. Rev. Lett.} \textbf{\bibinfo{volume}{62}},
  \bibinfo{pages}{2325} (\bibinfo{year}{1989}),
  \urlprefix\url{https://link.aps.org/doi/10.1103/PhysRevLett.62.2325}.

\bibitem[{\citenamefont{Peres}(1989)}]{PhysRevLett.62.2326}
\bibinfo{author}{\bibfnamefont{A.}~\bibnamefont{Peres}},
  \bibinfo{journal}{Phys. Rev. Lett.} \textbf{\bibinfo{volume}{62}},
  \bibinfo{pages}{2326} (\bibinfo{year}{1989}),
  \urlprefix\url{https://link.aps.org/doi/10.1103/PhysRevLett.62.2326}.

\bibitem[{\citenamefont{Li et~al.}(2013)\citenamefont{Li, Al-Amri, and
  Zubairy}}]{PhysRevA.88.046102}
\bibinfo{author}{\bibfnamefont{Z.-H.} \bibnamefont{Li}},
  \bibinfo{author}{\bibfnamefont{M.}~\bibnamefont{Al-Amri}}, \bibnamefont{and}
  \bibinfo{author}{\bibfnamefont{M.~S.} \bibnamefont{Zubairy}},
  \bibinfo{journal}{Phys. Rev. A} \textbf{\bibinfo{volume}{88}},
  \bibinfo{pages}{046102} (\bibinfo{year}{2013}),
  \urlprefix\url{https://link.aps.org/doi/10.1103/PhysRevA.88.046102}.

\bibitem[{\citenamefont{Bula et~al.}(2016)\citenamefont{Bula, Bartkiewicz,
  \ifmmode~\check{C}\else \v{C}\fi{}ernoch, Jav\ifmmode~\mathring{u}\else
  \r{u}\fi{}rek, Lemr, Mich\'alek, and Soubusta}}]{PhysRevA.94.052106}
\bibinfo{author}{\bibfnamefont{M.}~\bibnamefont{Bula}},
  \bibinfo{author}{\bibfnamefont{K.}~\bibnamefont{Bartkiewicz}},
  \bibinfo{author}{\bibfnamefont{A.}~\bibnamefont{\ifmmode~\check{C}\else
  \v{C}\fi{}ernoch}},
  \bibinfo{author}{\bibfnamefont{D.}~\bibnamefont{Jav\ifmmode~\mathring{u}\else
  \r{u}\fi{}rek}}, \bibinfo{author}{\bibfnamefont{K.}~\bibnamefont{Lemr}},
  \bibinfo{author}{\bibfnamefont{V.}~\bibnamefont{Mich\'alek}},
  \bibnamefont{and} \bibinfo{author}{\bibfnamefont{J.}~\bibnamefont{Soubusta}},
  \bibinfo{journal}{Phys. Rev. A} \textbf{\bibinfo{volume}{94}},
  \bibinfo{pages}{052106} (\bibinfo{year}{2016}),
  \urlprefix\url{https://link.aps.org/doi/10.1103/PhysRevA.94.052106}.

\bibitem[{\citenamefont{Sokolovski}(2017)}]{SOKOLOVSKI2017227}
\bibinfo{author}{\bibfnamefont{D.}~\bibnamefont{Sokolovski}},
  \bibinfo{journal}{Physics Letters A} \textbf{\bibinfo{volume}{381}},
  \bibinfo{pages}{227 } (\bibinfo{year}{2017}), ISSN \bibinfo{issn}{0375-9601},
  \urlprefix\url{http://www.sciencedirect.com/science/article/pii/S0375960116316590}.

\bibitem[{\citenamefont{Griffiths}(2016)}]{PhysRevA.94.032115}
\bibinfo{author}{\bibfnamefont{R.~B.} \bibnamefont{Griffiths}},
  \bibinfo{journal}{Phys. Rev. A} \textbf{\bibinfo{volume}{94}},
  \bibinfo{pages}{032115} (\bibinfo{year}{2016}),
  \urlprefix\url{https://link.aps.org/doi/10.1103/PhysRevA.94.032115}.

\bibitem[{\citenamefont{Alonso and Jordan}(2015)}]{Alonso2015}
\bibinfo{author}{\bibfnamefont{M.~A.} \bibnamefont{Alonso}} \bibnamefont{and}
  \bibinfo{author}{\bibfnamefont{A.~N.} \bibnamefont{Jordan}},
  \bibinfo{journal}{Quantum Studies: Mathematics and Foundations}
  \textbf{\bibinfo{volume}{2}}, \bibinfo{pages}{255} (\bibinfo{year}{2015}),
  ISSN \bibinfo{issn}{2196-5617},
  \urlprefix\url{https://doi.org/10.1007/s40509-015-0044-8}.

\bibitem[{\citenamefont{Poto\ifmmode~\check{c}\else \v{c}\fi{}ek and
  Ferenczi}(2015)}]{PhysRevA.92.023829}
\bibinfo{author}{\bibfnamefont{V.}~\bibnamefont{Poto\ifmmode~\check{c}\else
  \v{c}\fi{}ek}} \bibnamefont{and}
  \bibinfo{author}{\bibfnamefont{G.}~\bibnamefont{Ferenczi}},
  \bibinfo{journal}{Phys. Rev. A} \textbf{\bibinfo{volume}{92}},
  \bibinfo{pages}{023829} (\bibinfo{year}{2015}),
  \urlprefix\url{https://link.aps.org/doi/10.1103/PhysRevA.92.023829}.

\bibitem[{\citenamefont{Bartkiewicz et~al.}(2015)\citenamefont{Bartkiewicz,
  Černoch, Javůrek, Lemr, Soubusta, and Svozilík}}]{PhysRevA.91.012103}
\bibinfo{author}{\bibfnamefont{K.}~\bibnamefont{Bartkiewicz}},
  \bibinfo{author}{\bibfnamefont{A.}~\bibnamefont{Černoch}},
  \bibinfo{author}{\bibfnamefont{D.}~\bibnamefont{Javůrek}},
  \bibinfo{author}{\bibfnamefont{K.}~\bibnamefont{Lemr}},
  \bibinfo{author}{\bibfnamefont{J.}~\bibnamefont{Soubusta}}, \bibnamefont{and}
  \bibinfo{author}{\bibfnamefont{J.}~\bibnamefont{Svozilík}},
  \bibinfo{journal}{Phys. Rev. A} \textbf{\bibinfo{volume}{91}},
  \bibinfo{pages}{012103} (\bibinfo{year}{2015}),
  \urlprefix\url{https://link.aps.org/doi/10.1103/PhysRevA.91.012103}.

\bibitem[{\citenamefont{Saldanha}(2014)}]{PhysRevA.89.033825}
\bibinfo{author}{\bibfnamefont{P.~L.} \bibnamefont{Saldanha}},
  \bibinfo{journal}{Phys. Rev. A} \textbf{\bibinfo{volume}{89}},
  \bibinfo{pages}{033825} (\bibinfo{year}{2014}),
  \urlprefix\url{https://link.aps.org/doi/10.1103/PhysRevA.89.033825}.

\bibitem[{\citenamefont{Vaidman}(2013)}]{PhysRevA.88.046103}
\bibinfo{author}{\bibfnamefont{L.}~\bibnamefont{Vaidman}},
  \bibinfo{journal}{Phys. Rev. A} \textbf{\bibinfo{volume}{88}},
  \bibinfo{pages}{046103} (\bibinfo{year}{2013}),
  \urlprefix\url{https://link.aps.org/doi/10.1103/PhysRevA.88.046103}.

\bibitem[{\citenamefont{Danan et~al.}(2013)\citenamefont{Danan, Farfurnik,
  Bar-Ad, and Vaidman}}]{PhysRevLett.111.240402}
\bibinfo{author}{\bibfnamefont{A.}~\bibnamefont{Danan}},
  \bibinfo{author}{\bibfnamefont{D.}~\bibnamefont{Farfurnik}},
  \bibinfo{author}{\bibfnamefont{S.}~\bibnamefont{Bar-Ad}}, \bibnamefont{and}
  \bibinfo{author}{\bibfnamefont{L.}~\bibnamefont{Vaidman}},
  \bibinfo{journal}{Phys. Rev. Lett.} \textbf{\bibinfo{volume}{111}},
  \bibinfo{pages}{240402} (\bibinfo{year}{2013}),
  \urlprefix\url{https://link.aps.org/doi/10.1103/PhysRevLett.111.240402}.

\bibitem[{\citenamefont{Griffiths}(2018)}]{PhysRevA.97.026102}
\bibinfo{author}{\bibfnamefont{R.~B.} \bibnamefont{Griffiths}},
  \bibinfo{journal}{Phys. Rev. A} \textbf{\bibinfo{volume}{97}},
  \bibinfo{pages}{026102} (\bibinfo{year}{2018}),
  \urlprefix\url{https://link.aps.org/doi/10.1103/PhysRevA.97.026102}.

\bibitem[{\citenamefont{Vaidman}(2017{\natexlab{a}})}]{PhysRevA.95.066101}
\bibinfo{author}{\bibfnamefont{L.}~\bibnamefont{Vaidman}},
  \bibinfo{journal}{Phys. Rev. A} \textbf{\bibinfo{volume}{95}},
  \bibinfo{pages}{066101} (\bibinfo{year}{2017}{\natexlab{a}}),
  \urlprefix\url{https://link.aps.org/doi/10.1103/PhysRevA.95.066101}.

\bibitem[{\citenamefont{Vaidman}(2017{\natexlab{b}})}]{vaidman2017comment}
\bibinfo{author}{\bibfnamefont{L.}~\bibnamefont{Vaidman}},
  \bibinfo{journal}{arXiv preprint arXiv:1703.03615}
  (\bibinfo{year}{2017}{\natexlab{b}}).

\bibitem[{\citenamefont{Vaidman}(2018)}]{Vaidman_2018a}
\bibinfo{author}{\bibfnamefont{L.}~\bibnamefont{Vaidman}},
  \bibinfo{journal}{Journal of Physics A: Mathematical and Theoretical}
  \textbf{\bibinfo{volume}{51}}, \bibinfo{pages}{068002}
  (\bibinfo{year}{2018}),
  \urlprefix\url{https://doi.org/10.1088%2F1751-8121%2Faa8d24}.

\bibitem[{\citenamefont{Aharonov and Vaidman}(1989)}]{PhysRevLett.62.2327}
\bibinfo{author}{\bibfnamefont{Y.}~\bibnamefont{Aharonov}} \bibnamefont{and}
  \bibinfo{author}{\bibfnamefont{L.}~\bibnamefont{Vaidman}},
  \bibinfo{journal}{Phys. Rev. Lett.} \textbf{\bibinfo{volume}{62}},
  \bibinfo{pages}{2327} (\bibinfo{year}{1989}),
  \urlprefix\url{https://link.aps.org/doi/10.1103/PhysRevLett.62.2327}.

\bibitem[{\citenamefont{Vaidman}(2017{\natexlab{c}})}]{doi:10.1098/rsta.2016.0395}
\bibinfo{author}{\bibfnamefont{L.}~\bibnamefont{Vaidman}},
  \bibinfo{journal}{Philosophical Transactions of the Royal Society A:
  Mathematical, Physical and Engineering Sciences}
  \textbf{\bibinfo{volume}{375}}, \bibinfo{pages}{20160395}
  (\bibinfo{year}{2017}{\natexlab{c}}),
  \urlprefix\url{https://royalsocietypublishing.org/doi/abs/10.1098/rsta.2016.0395}.

\bibitem[{\citenamefont{Bartkiewicz et~al.}(2016)\citenamefont{Bartkiewicz,
  \ifmmode~\check{C}\else \v{C}\fi{}ernoch, Jav\ifmmode~\mathring{u}\else
  \r{u}\fi{}rek, Lemr, Soubusta, and Svozil\'{\i}k}}]{PhysRevA.93.036104}
\bibinfo{author}{\bibfnamefont{K.}~\bibnamefont{Bartkiewicz}},
  \bibinfo{author}{\bibfnamefont{A.}~\bibnamefont{\ifmmode~\check{C}\else
  \v{C}\fi{}ernoch}},
  \bibinfo{author}{\bibfnamefont{D.}~\bibnamefont{Jav\ifmmode~\mathring{u}\else
  \r{u}\fi{}rek}}, \bibinfo{author}{\bibfnamefont{K.}~\bibnamefont{Lemr}},
  \bibinfo{author}{\bibfnamefont{J.}~\bibnamefont{Soubusta}}, \bibnamefont{and}
  \bibinfo{author}{\bibfnamefont{J.~c.~v.} \bibnamefont{Svozil\'{\i}k}},
  \bibinfo{journal}{Phys. Rev. A} \textbf{\bibinfo{volume}{93}},
  \bibinfo{pages}{036104} (\bibinfo{year}{2016}),
  \urlprefix\url{https://link.aps.org/doi/10.1103/PhysRevA.93.036104}.

\bibitem[{\citenamefont{Vaidman}(2016{\natexlab{a}})}]{PhysRevA.93.036103}
\bibinfo{author}{\bibfnamefont{L.}~\bibnamefont{Vaidman}},
  \bibinfo{journal}{Phys. Rev. A} \textbf{\bibinfo{volume}{93}},
  \bibinfo{pages}{036103} (\bibinfo{year}{2016}{\natexlab{a}}),
  \urlprefix\url{https://link.aps.org/doi/10.1103/PhysRevA.93.036103}.

\bibitem[{\citenamefont{Vaidman}(2016{\natexlab{b}})}]{PhysRevA.93.017801}
\bibinfo{author}{\bibfnamefont{L.}~\bibnamefont{Vaidman}},
  \bibinfo{journal}{Phys. Rev. A} \textbf{\bibinfo{volume}{93}},
  \bibinfo{pages}{017801} (\bibinfo{year}{2016}{\natexlab{b}}),
  \urlprefix\url{https://link.aps.org/doi/10.1103/PhysRevA.93.017801}.

\bibitem[{\citenamefont{Vaidman and Tsutsui}(2018)}]{Vaidman_2018}
\bibinfo{author}{\bibfnamefont{L.}~\bibnamefont{Vaidman}} \bibnamefont{and}
  \bibinfo{author}{\bibfnamefont{I.}~\bibnamefont{Tsutsui}},
  \bibinfo{journal}{Entropy} \textbf{\bibinfo{volume}{20}},
  \bibinfo{pages}{538} (\bibinfo{year}{2018}), ISSN \bibinfo{issn}{1099-4300},
  \urlprefix\url{http://dx.doi.org/10.3390/e20070538}.

\bibitem[{\citenamefont{Vaidman}(2017{\natexlab{d}})}]{vaidman2017comments}
\bibinfo{author}{\bibfnamefont{L.}~\bibnamefont{Vaidman}},
  \bibinfo{journal}{arXiv preprint arXiv:1703.01616}
  (\bibinfo{year}{2017}{\natexlab{d}}).

\bibitem[{\citenamefont{Xu et~al.}(2019)\citenamefont{Xu, Pan, Wang, Dziewior,
  Knips, Kedem, Sun, Xu, Han, Li et~al.}}]{PhysRevLett.122.100405}
\bibinfo{author}{\bibfnamefont{X.-Y.} \bibnamefont{Xu}},
  \bibinfo{author}{\bibfnamefont{W.-W.} \bibnamefont{Pan}},
  \bibinfo{author}{\bibfnamefont{Q.-Q.} \bibnamefont{Wang}},
  \bibinfo{author}{\bibfnamefont{J.}~\bibnamefont{Dziewior}},
  \bibinfo{author}{\bibfnamefont{L.}~\bibnamefont{Knips}},
  \bibinfo{author}{\bibfnamefont{Y.}~\bibnamefont{Kedem}},
  \bibinfo{author}{\bibfnamefont{K.}~\bibnamefont{Sun}},
  \bibinfo{author}{\bibfnamefont{J.-S.} \bibnamefont{Xu}},
  \bibinfo{author}{\bibfnamefont{Y.-J.} \bibnamefont{Han}},
  \bibinfo{author}{\bibfnamefont{C.-F.} \bibnamefont{Li}},
  \bibnamefont{et~al.}, \bibinfo{journal}{Phys. Rev. Lett.}
  \textbf{\bibinfo{volume}{122}}, \bibinfo{pages}{100405}
  (\bibinfo{year}{2019}),
  \urlprefix\url{https://link.aps.org/doi/10.1103/PhysRevLett.122.100405}.

\bibitem[{\citenamefont{Peleg and Vaidman}(2019)}]{PhysRevA.99.026103}
\bibinfo{author}{\bibfnamefont{U.}~\bibnamefont{Peleg}} \bibnamefont{and}
  \bibinfo{author}{\bibfnamefont{L.}~\bibnamefont{Vaidman}},
  \bibinfo{journal}{Phys. Rev. A} \textbf{\bibinfo{volume}{99}},
  \bibinfo{pages}{026103} (\bibinfo{year}{2019}),
  \urlprefix\url{https://link.aps.org/doi/10.1103/PhysRevA.99.026103}.

\bibitem[{\citenamefont{Pryde et~al.}(2004)\citenamefont{Pryde, O'Brien, White,
  Bartlett, and Ralph}}]{PhysRevLett.92.190402}
\bibinfo{author}{\bibfnamefont{G.~J.} \bibnamefont{Pryde}},
  \bibinfo{author}{\bibfnamefont{J.~L.} \bibnamefont{O'Brien}},
  \bibinfo{author}{\bibfnamefont{A.~G.} \bibnamefont{White}},
  \bibinfo{author}{\bibfnamefont{S.~D.} \bibnamefont{Bartlett}},
  \bibnamefont{and} \bibinfo{author}{\bibfnamefont{T.~C.} \bibnamefont{Ralph}},
  \bibinfo{journal}{Phys. Rev. Lett.} \textbf{\bibinfo{volume}{92}},
  \bibinfo{pages}{190402} (\bibinfo{year}{2004}),
  \urlprefix\url{https://link.aps.org/doi/10.1103/PhysRevLett.92.190402}.

\bibitem[{\citenamefont{Kim et~al.}(2018)\citenamefont{Kim, Kim, Lee, Han,
  Moon, Kim, and Cho}}]{Kim2018}
\bibinfo{author}{\bibfnamefont{Y.}~\bibnamefont{Kim}},
  \bibinfo{author}{\bibfnamefont{Y.-S.} \bibnamefont{Kim}},
  \bibinfo{author}{\bibfnamefont{S.-Y.} \bibnamefont{Lee}},
  \bibinfo{author}{\bibfnamefont{S.-W.} \bibnamefont{Han}},
  \bibinfo{author}{\bibfnamefont{S.}~\bibnamefont{Moon}},
  \bibinfo{author}{\bibfnamefont{Y.-H.} \bibnamefont{Kim}}, \bibnamefont{and}
  \bibinfo{author}{\bibfnamefont{Y.-W.} \bibnamefont{Cho}},
  \bibinfo{journal}{Nature Communications} \textbf{\bibinfo{volume}{9}},
  \bibinfo{pages}{192} (\bibinfo{year}{2018}), ISSN \bibinfo{issn}{2041-1723},
  \urlprefix\url{https://doi.org/10.1038/s41467-017-02511-2}.

\bibitem[{\citenamefont{Lundeen et~al.}(2011)\citenamefont{Lundeen, Sutherland,
  Patel, Stewart, and Bamber}}]{Lundeen2011}
\bibinfo{author}{\bibfnamefont{J.~S.} \bibnamefont{Lundeen}},
  \bibinfo{author}{\bibfnamefont{B.}~\bibnamefont{Sutherland}},
  \bibinfo{author}{\bibfnamefont{A.}~\bibnamefont{Patel}},
  \bibinfo{author}{\bibfnamefont{C.}~\bibnamefont{Stewart}}, \bibnamefont{and}
  \bibinfo{author}{\bibfnamefont{C.}~\bibnamefont{Bamber}},
  \bibinfo{journal}{Nature} \textbf{\bibinfo{volume}{474}}, \bibinfo{pages}{188
  EP } (\bibinfo{year}{2011}),
  \urlprefix\url{https://doi.org/10.1038/nature10120}.

\bibitem[{\citenamefont{Matzkin}(2012)}]{PhysRevLett.109.150407}
\bibinfo{author}{\bibfnamefont{A.}~\bibnamefont{Matzkin}},
  \bibinfo{journal}{Phys. Rev. Lett.} \textbf{\bibinfo{volume}{109}},
  \bibinfo{pages}{150407} (\bibinfo{year}{2012}),
  \urlprefix\url{https://link.aps.org/doi/10.1103/PhysRevLett.109.150407}.

\bibitem[{\citenamefont{Matzkin and Pan}(2013)}]{1751-8121-46-31-315307}
\bibinfo{author}{\bibfnamefont{A.}~\bibnamefont{Matzkin}} \bibnamefont{and}
  \bibinfo{author}{\bibfnamefont{A.~K.} \bibnamefont{Pan}},
  \bibinfo{journal}{Journal of Physics A: Mathematical and Theoretical}
  \textbf{\bibinfo{volume}{46}}, \bibinfo{pages}{315307}
  (\bibinfo{year}{2013}),
  \urlprefix\url{http://stacks.iop.org/1751-8121/46/i=31/a=315307}.

\bibitem[{\citenamefont{Hallaji et~al.}(2017)\citenamefont{Hallaji, Feizpour,
  Dmochowski, Sinclair, and Steinberg}}]{Hallaji2017}
\bibinfo{author}{\bibfnamefont{M.}~\bibnamefont{Hallaji}},
  \bibinfo{author}{\bibfnamefont{A.}~\bibnamefont{Feizpour}},
  \bibinfo{author}{\bibfnamefont{G.}~\bibnamefont{Dmochowski}},
  \bibinfo{author}{\bibfnamefont{J.}~\bibnamefont{Sinclair}}, \bibnamefont{and}
  \bibinfo{author}{\bibfnamefont{A.}~\bibnamefont{Steinberg}},
  \bibinfo{journal}{Nature Physics} \textbf{\bibinfo{volume}{13}},
  \bibinfo{pages}{540 EP } (\bibinfo{year}{2017}),
  \urlprefix\url{https://doi.org/10.1038/nphys4040}.

\bibitem[{\citenamefont{Ahnert and Payne}(2004)}]{PhysRevA.70.042102}
\bibinfo{author}{\bibfnamefont{S.~E.} \bibnamefont{Ahnert}} \bibnamefont{and}
  \bibinfo{author}{\bibfnamefont{M.~C.} \bibnamefont{Payne}},
  \bibinfo{journal}{Phys. Rev. A} \textbf{\bibinfo{volume}{70}},
  \bibinfo{pages}{042102} (\bibinfo{year}{2004}),
  \urlprefix\url{https://link.aps.org/doi/10.1103/PhysRevA.70.042102}.

\bibitem[{\citenamefont{Aharonov et~al.}(2002)\citenamefont{Aharonov, Botero,
  Popescu, Reznik, and Tollaksen}}]{AHARONOV2002130}
\bibinfo{author}{\bibfnamefont{Y.}~\bibnamefont{Aharonov}},
  \bibinfo{author}{\bibfnamefont{A.}~\bibnamefont{Botero}},
  \bibinfo{author}{\bibfnamefont{S.}~\bibnamefont{Popescu}},
  \bibinfo{author}{\bibfnamefont{B.}~\bibnamefont{Reznik}}, \bibnamefont{and}
  \bibinfo{author}{\bibfnamefont{J.}~\bibnamefont{Tollaksen}},
  \bibinfo{journal}{Physics Letters A} \textbf{\bibinfo{volume}{301}},
  \bibinfo{pages}{130 } (\bibinfo{year}{2002}), ISSN \bibinfo{issn}{0375-9601},
  \urlprefix\url{http://www.sciencedirect.com/science/article/pii/S0375960102009866}.

\bibitem[{\citenamefont{Pryde et~al.}(2005)\citenamefont{Pryde, O'Brien, White,
  Ralph, and Wiseman}}]{PhysRevLett.94.220405}
\bibinfo{author}{\bibfnamefont{G.~J.} \bibnamefont{Pryde}},
  \bibinfo{author}{\bibfnamefont{J.~L.} \bibnamefont{O'Brien}},
  \bibinfo{author}{\bibfnamefont{A.~G.} \bibnamefont{White}},
  \bibinfo{author}{\bibfnamefont{T.~C.} \bibnamefont{Ralph}}, \bibnamefont{and}
  \bibinfo{author}{\bibfnamefont{H.~M.} \bibnamefont{Wiseman}},
  \bibinfo{journal}{Phys. Rev. Lett.} \textbf{\bibinfo{volume}{94}},
  \bibinfo{pages}{220405} (\bibinfo{year}{2005}),
  \urlprefix\url{https://link.aps.org/doi/10.1103/PhysRevLett.94.220405}.

\bibitem[{\citenamefont{Johansen}(2007)}]{JOHANSEN2007374}
\bibinfo{author}{\bibfnamefont{L.~M.} \bibnamefont{Johansen}},
  \bibinfo{journal}{Physics Letters A} \textbf{\bibinfo{volume}{366}},
  \bibinfo{pages}{374 } (\bibinfo{year}{2007}), ISSN \bibinfo{issn}{0375-9601},
  \urlprefix\url{http://www.sciencedirect.com/science/article/pii/S0375960107002654}.

\bibitem[{\citenamefont{Dressel et~al.}(2014)\citenamefont{Dressel, Malik,
  Miatto, Jordan, and Boyd}}]{RevModPhys.86.307}
\bibinfo{author}{\bibfnamefont{J.}~\bibnamefont{Dressel}},
  \bibinfo{author}{\bibfnamefont{M.}~\bibnamefont{Malik}},
  \bibinfo{author}{\bibfnamefont{F.~M.} \bibnamefont{Miatto}},
  \bibinfo{author}{\bibfnamefont{A.~N.} \bibnamefont{Jordan}},
  \bibnamefont{and} \bibinfo{author}{\bibfnamefont{R.~W.} \bibnamefont{Boyd}},
  \bibinfo{journal}{Rev. Mod. Phys.} \textbf{\bibinfo{volume}{86}},
  \bibinfo{pages}{307} (\bibinfo{year}{2014}),
  \urlprefix\url{https://link.aps.org/doi/10.1103/RevModPhys.86.307}.

\bibitem[{\citenamefont{Cohen and Pollak}(2018)}]{PhysRevA.98.042112}
\bibinfo{author}{\bibfnamefont{E.}~\bibnamefont{Cohen}} \bibnamefont{and}
  \bibinfo{author}{\bibfnamefont{E.}~\bibnamefont{Pollak}},
  \bibinfo{journal}{Phys. Rev. A} \textbf{\bibinfo{volume}{98}},
  \bibinfo{pages}{042112} (\bibinfo{year}{2018}),
  \urlprefix\url{https://link.aps.org/doi/10.1103/PhysRevA.98.042112}.

\bibitem[{\citenamefont{Denkmayr et~al.}(2018)\citenamefont{Denkmayr, Dressel,
  Geppert-Kleinrath, Hasegawa, and Sponar}}]{DENKMAYR2018339}
\bibinfo{author}{\bibfnamefont{T.}~\bibnamefont{Denkmayr}},
  \bibinfo{author}{\bibfnamefont{J.}~\bibnamefont{Dressel}},
  \bibinfo{author}{\bibfnamefont{H.}~\bibnamefont{Geppert-Kleinrath}},
  \bibinfo{author}{\bibfnamefont{Y.}~\bibnamefont{Hasegawa}}, \bibnamefont{and}
  \bibinfo{author}{\bibfnamefont{S.}~\bibnamefont{Sponar}},
  \bibinfo{journal}{Physica B: Condensed Matter}
  \textbf{\bibinfo{volume}{551}}, \bibinfo{pages}{339 } (\bibinfo{year}{2018}),
  ISSN \bibinfo{issn}{0921-4526}, \bibinfo{note}{the 11th International
  Conference on Neutron Scattering (ICNS 2017)},
  \urlprefix\url{http://www.sciencedirect.com/science/article/pii/S0921452618302722}.

\bibitem[{\citenamefont{Vallone and Dequal}(2016)}]{PhysRevLett.116.040502}
\bibinfo{author}{\bibfnamefont{G.}~\bibnamefont{Vallone}} \bibnamefont{and}
  \bibinfo{author}{\bibfnamefont{D.}~\bibnamefont{Dequal}},
  \bibinfo{journal}{Phys. Rev. Lett.} \textbf{\bibinfo{volume}{116}},
  \bibinfo{pages}{040502} (\bibinfo{year}{2016}),
  \urlprefix\url{https://link.aps.org/doi/10.1103/PhysRevLett.116.040502}.

\bibitem[{\citenamefont{Dziewior et~al.}(2019)\citenamefont{Dziewior, Knips,
  Farfurnik, Senkalla, Benshalom, Efroni, Meinecke, Bar-Ad, Weinfurter, and
  Vaidman}}]{Dziewior2881}
\bibinfo{author}{\bibfnamefont{J.}~\bibnamefont{Dziewior}},
  \bibinfo{author}{\bibfnamefont{L.}~\bibnamefont{Knips}},
  \bibinfo{author}{\bibfnamefont{D.}~\bibnamefont{Farfurnik}},
  \bibinfo{author}{\bibfnamefont{K.}~\bibnamefont{Senkalla}},
  \bibinfo{author}{\bibfnamefont{N.}~\bibnamefont{Benshalom}},
  \bibinfo{author}{\bibfnamefont{J.}~\bibnamefont{Efroni}},
  \bibinfo{author}{\bibfnamefont{J.}~\bibnamefont{Meinecke}},
  \bibinfo{author}{\bibfnamefont{S.}~\bibnamefont{Bar-Ad}},
  \bibinfo{author}{\bibfnamefont{H.}~\bibnamefont{Weinfurter}},
  \bibnamefont{and} \bibinfo{author}{\bibfnamefont{L.}~\bibnamefont{Vaidman}},
  \bibinfo{journal}{Proceedings of the National Academy of Sciences}
  \textbf{\bibinfo{volume}{116}}, \bibinfo{pages}{2881} (\bibinfo{year}{2019}),
  ISSN \bibinfo{issn}{0027-8424},
  \eprint{https://www.pnas.org/content/116/8/2881.full.pdf},
  \urlprefix\url{https://www.pnas.org/content/116/8/2881}.

\bibitem[{\citenamefont{Zhang et~al.}(2016)\citenamefont{Zhang, Wu, and
  Chen}}]{PhysRevA.93.032128}
\bibinfo{author}{\bibfnamefont{Y.-X.} \bibnamefont{Zhang}},
  \bibinfo{author}{\bibfnamefont{S.}~\bibnamefont{Wu}}, \bibnamefont{and}
  \bibinfo{author}{\bibfnamefont{Z.-B.} \bibnamefont{Chen}},
  \bibinfo{journal}{Phys. Rev. A} \textbf{\bibinfo{volume}{93}},
  \bibinfo{pages}{032128} (\bibinfo{year}{2016}),
  \urlprefix\url{https://link.aps.org/doi/10.1103/PhysRevA.93.032128}.

\bibitem[{\citenamefont{Zhu et~al.}(2016)\citenamefont{Zhu, Zhang, and
  Wu}}]{PhysRevA.93.062304}
\bibinfo{author}{\bibfnamefont{X.}~\bibnamefont{Zhu}},
  \bibinfo{author}{\bibfnamefont{Y.-X.} \bibnamefont{Zhang}}, \bibnamefont{and}
  \bibinfo{author}{\bibfnamefont{S.}~\bibnamefont{Wu}}, \bibinfo{journal}{Phys.
  Rev. A} \textbf{\bibinfo{volume}{93}}, \bibinfo{pages}{062304}
  (\bibinfo{year}{2016}),
  \urlprefix\url{https://link.aps.org/doi/10.1103/PhysRevA.93.062304}.

\bibitem[{\citenamefont{Zou et~al.}(2015)\citenamefont{Zou, Zhang, and
  Song}}]{PhysRevA.91.052109}
\bibinfo{author}{\bibfnamefont{P.}~\bibnamefont{Zou}},
  \bibinfo{author}{\bibfnamefont{Z.-M.} \bibnamefont{Zhang}}, \bibnamefont{and}
  \bibinfo{author}{\bibfnamefont{W.}~\bibnamefont{Song}},
  \bibinfo{journal}{Phys. Rev. A} \textbf{\bibinfo{volume}{91}},
  \bibinfo{pages}{052109} (\bibinfo{year}{2015}),
  \urlprefix\url{https://link.aps.org/doi/10.1103/PhysRevA.91.052109}.

\bibitem[{\citenamefont{Brodutch and Cohen}(2016)}]{PhysRevLett.116.070404}
\bibinfo{author}{\bibfnamefont{A.}~\bibnamefont{Brodutch}} \bibnamefont{and}
  \bibinfo{author}{\bibfnamefont{E.}~\bibnamefont{Cohen}},
  \bibinfo{journal}{Phys. Rev. Lett.} \textbf{\bibinfo{volume}{116}},
  \bibinfo{pages}{070404} (\bibinfo{year}{2016}),
  \urlprefix\url{https://link.aps.org/doi/10.1103/PhysRevLett.116.070404}.

\bibitem[{\citenamefont{Lanyon et~al.}(2008)\citenamefont{Lanyon, Barbieri,
  Almeida, Jennewein, Ralph, Resch, Pryde, O'Brien, Gilchrist, and
  White}}]{Lanyon2008}
\bibinfo{author}{\bibfnamefont{B.~P.} \bibnamefont{Lanyon}},
  \bibinfo{author}{\bibfnamefont{M.}~\bibnamefont{Barbieri}},
  \bibinfo{author}{\bibfnamefont{M.~P.} \bibnamefont{Almeida}},
  \bibinfo{author}{\bibfnamefont{T.}~\bibnamefont{Jennewein}},
  \bibinfo{author}{\bibfnamefont{T.~C.} \bibnamefont{Ralph}},
  \bibinfo{author}{\bibfnamefont{K.~J.} \bibnamefont{Resch}},
  \bibinfo{author}{\bibfnamefont{G.~J.} \bibnamefont{Pryde}},
  \bibinfo{author}{\bibfnamefont{J.~L.} \bibnamefont{O'Brien}},
  \bibinfo{author}{\bibfnamefont{A.}~\bibnamefont{Gilchrist}},
  \bibnamefont{and} \bibinfo{author}{\bibfnamefont{A.~G.} \bibnamefont{White}},
  \bibinfo{journal}{Nature Physics} \textbf{\bibinfo{volume}{5}},
  \bibinfo{pages}{134 EP } (\bibinfo{year}{2008}), \bibinfo{note}{article},
  \urlprefix\url{https://doi.org/10.1038/nphys1150}.

\bibitem[{\citenamefont{Lemr et~al.}(2011)\citenamefont{Lemr,
  \ifmmode~\check{C}\else \v{C}\fi{}ernoch, Soubusta, Kieling, Eisert, and
  Du\ifmmode~\check{s}\else \v{s}\fi{}ek}}]{PhysRevLett.106.013602}
\bibinfo{author}{\bibfnamefont{K.}~\bibnamefont{Lemr}},
  \bibinfo{author}{\bibfnamefont{A.}~\bibnamefont{\ifmmode~\check{C}\else
  \v{C}\fi{}ernoch}},
  \bibinfo{author}{\bibfnamefont{J.}~\bibnamefont{Soubusta}},
  \bibinfo{author}{\bibfnamefont{K.}~\bibnamefont{Kieling}},
  \bibinfo{author}{\bibfnamefont{J.}~\bibnamefont{Eisert}}, \bibnamefont{and}
  \bibinfo{author}{\bibfnamefont{M.}~\bibnamefont{Du\ifmmode~\check{s}\else
  \v{s}\fi{}ek}}, \bibinfo{journal}{Phys. Rev. Lett.}
  \textbf{\bibinfo{volume}{106}}, \bibinfo{pages}{013602}
  (\bibinfo{year}{2011}),
  \urlprefix\url{https://link.aps.org/doi/10.1103/PhysRevLett.106.013602}.

\bibitem[{\citenamefont{Kiesel et~al.}(2005)\citenamefont{Kiesel, Schmid,
  Weber, Ursin, and Weinfurter}}]{PhysRevLett.95.210505}
\bibinfo{author}{\bibfnamefont{N.}~\bibnamefont{Kiesel}},
  \bibinfo{author}{\bibfnamefont{C.}~\bibnamefont{Schmid}},
  \bibinfo{author}{\bibfnamefont{U.}~\bibnamefont{Weber}},
  \bibinfo{author}{\bibfnamefont{R.}~\bibnamefont{Ursin}}, \bibnamefont{and}
  \bibinfo{author}{\bibfnamefont{H.}~\bibnamefont{Weinfurter}},
  \bibinfo{journal}{Phys. Rev. Lett.} \textbf{\bibinfo{volume}{95}},
  \bibinfo{pages}{210505} (\bibinfo{year}{2005}),
  \urlprefix\url{https://link.aps.org/doi/10.1103/PhysRevLett.95.210505}.

\bibitem[{\citenamefont{Lemr et~al.}(2013)\citenamefont{Lemr, Bartkiewicz,
  \ifmmode~\check{C}\else \v{C}\fi{}ernoch, and Soubusta}}]{PhysRevA.87.062333}
\bibinfo{author}{\bibfnamefont{K.}~\bibnamefont{Lemr}},
  \bibinfo{author}{\bibfnamefont{K.}~\bibnamefont{Bartkiewicz}},
  \bibinfo{author}{\bibfnamefont{A.}~\bibnamefont{\ifmmode~\check{C}\else
  \v{C}\fi{}ernoch}}, \bibnamefont{and}
  \bibinfo{author}{\bibfnamefont{J.}~\bibnamefont{Soubusta}},
  \bibinfo{journal}{Phys. Rev. A} \textbf{\bibinfo{volume}{87}},
  \bibinfo{pages}{062333} (\bibinfo{year}{2013}),
  \urlprefix\url{https://link.aps.org/doi/10.1103/PhysRevA.87.062333}.

\end{thebibliography}
\end{document}